\documentclass[twocolumn,showpacs,preprintnumbers,prb]{revtex4}
\usepackage{graphicx,bm,amsmath,amssymb}

\def\gz{\ifmmode{Z\hskip -4.8pt Z}
    \else{\hbox{$Z\hskip -4.8pt Z$}}\fi}

\newcommand{\be}{\begin{equation}}
\newcommand{\ee}{\end{equation}}
\newcommand{\bea}{\begin{eqnarray}}
\newcommand{\eea}{\end{eqnarray}}

\begin{document}

\title{Universal scaling in nonequilibrium transport through an Anderson impurity}
\author{Juli\'{a}n~Rinc\'{o}n}
\author{A.~A.~Aligia}
\email{aligia@cab.cnea.gov.ar}
\author{K.~Hallberg}
\affiliation{Centro At\'{o}mico Bariloche and Instituto Balseiro, Comisi\'{o}n Nacional
de Energ\'{\i}a At\'{o}mica, 8400 Bariloche, Argentina}
\date{\today}
\date{\today}

\begin{abstract}
Using non-equilibrium renormalized perturbation theory, we calculate the
conductance $G$ as a function of temperature $T$ and bias voltage $V$ for an
Anderson model, suitable for describing transport properties through a
quantum dot. For renormalized parameters that correspond to the extreme
Kondo limit, we do not find a simple scaling formula beyond a quadratic dependence in 
$T$ and $V$. However, if valence fluctuations are allowed, we find excellent
agreement with recent experiments.
\end{abstract}

\maketitle

Universality is one of the most beautiful and useful concepts in physics. In
general, the physical properties of a system depend on a certain number of
parameters which change for different experimental realizations. However, in
favorable cases, physical observables are described by the same universal
function, once the different physical magnitudes are scaled appropriately.
For example in the case of the temperature dependence of the conductance
through one quantum dot $G(T)$ in the limit of zero bias voltage $V$, once a
characteristic energy scale $T_{K}$ (the Kondo temperature) is identified,
the conductance of different systems is very well described by the same
universal function $G(T/T_{K})$, even if the systems have very different $%
T_{K}$.\cite{gold,grobis} Scaling and universality are concepts which are
quite naturally connected to renormalization group treatments of the
Anderson model in the Kondo regime (Coulomb repulsion $U$ much
larger than the resonant level width $\Delta $), and in fact numerical
renormalization group (NRG) calculations reproduce the scaling mentioned
above and in other physical properties.\cite{bulla,chz}

Theoretically, the situation is much more difficult in the nonequilibrium
situation which arises for a finite bias voltage between the leads connected
to the quantum dot in the experiment. Only recently, extensions to the
nonequilibrium case of essentially exact techniques such as NRG \cite{anders}
and exact Bethe ansatz \cite{mehta} were proposed, while approximations used
at equilibrium have shortcomings when extended to the nonequilibrium case.
\cite{none} Nevertheless, using a Fermi liquid approach, based on
perturbation theory (PT) in $U/\Delta $, and Ward identities, Oguri has
determined exactly the scaling for $T$ and $eV$ small compared to $T_{K}$
for the Anderson model \cite{ogu}
\begin{equation}
G(T,V)=G_{0}\left[ 1-c_{T}\left( \frac{T}{T_{K}}\right) ^{2}-\alpha
\,c_{T}\left( \frac{eV}{kT_{K}}\right) ^{2}+\dots \right] ,  \label{g1}
\end{equation}
where $G_{0}=G(0,0)$ and the values of $c_{T}$ and $\alpha $ are discussed
below.

Recent experiments in GaAs quantum dots for different situations in
the nonequilibrium regime,\cite{grobis} for low $T$ and $V$ have found that 
$G(T,V)$ is well described by a universal scaling function that extends 
Eq. (\ref{g1}) to higher temperatures
\begin{equation}
\frac{G(T,V)}{G_{E}(T)}\simeq 1-\frac{\alpha \,c_{T}(eV/kT_{K})^{2}}{%
1+(\gamma /\alpha -1)\,c_{T}\,(T/T_{K})^{2}}.  \label{g2}
\end{equation}
Here $c_{T}\simeq 5.488$ is fixed by Eqs. (\ref{g1}) and (\ref{ge}), $\alpha
=0.10\pm 0.015$, $\gamma =0.5\pm 0.1$ and $G_{E}(T)$ is an empirical curve
obtained from a fit to NRG results:
\begin{equation}
G_{E}(T)=\frac{G_{0}}{\left[ 1+(2^{1/s}-1)(T/T_{K})^{2}\right] ^{s}},
\label{ge}
\end{equation}
with $s=0.21$ for an impurity with total spin $S=1/2$. From these
equations, one can see that $\alpha $ is the ratio of the term of
order $[eV/(kT_{K})]^{2}$\ with respect to that of order 
$\left(T/T_{K}\right) ^{2}$ in the decrease in the conductance, while 
$\gamma $ represents the effect of terms $[eV/(kT_{K})]^{2}\left(
T/T_{K}\right) ^{2n}$ with integer $n>1$.

From 
%the low $T$ and $V$ expansion of the conductance of 
an exactly solvable
anisotropic Kondo model,\cite{schi} one extracts $\alpha =3/\pi ^{2}\simeq
0.30$ and $\gamma =2(\pi T_{K}/T_{a})^{2}/c_{T}\simeq 3.60(T_{K}/T_{a})^{2}$, 
where $T_{a}$ is an energy scale of the order of $T_{K}$. To our
knowledge, no other precise information on $\gamma $ exists.

The purpose of this work is to test the observed scaling relation and
calculate $\gamma $ in the impurity Anderson model, using renormalized PT
(RPT).\cite{he1} The basic idea of RPT is to reorganize the PT in terms of
fully dressed quasiparticles in a Fermi liquid picture. The main
advantage is that even in the strong coupling (SC) limit $U\rightarrow
\infty $, for which ordinary PT in $u=U/(\pi \Delta )$
becomes invalid, the corresponding ratio between renormalized parameters
(denoted by a tilde) becomes $\widetilde{u}\equiv \widetilde{U}/(\pi 
\widetilde{\Delta })\rightarrow 1$, being $\widetilde{u}<1$ 
for finite $U$.\cite{he1} For nontrivial cases, already free
quasiparticles (taking $\widetilde{U}=0$, which is similar to slave
bosons in a mean field approximation \cite{vaug}) reproduce very well the
low-frequency part of the equilibrium spectral density at the quantum dot.
An example is a case in which the Kondo peak is split in two.\cite{vaug} 
$\widetilde{U}$ (proportional to the full vertex) represents the
\textquotedblleft residual\textquotedblright\ interaction between
quasiparticles. Calculating the renormalized retarded self-energy 
$\widetilde{\Sigma }^{r}$ within nonequilibrium RPT up to second order in 
$\widetilde{u}$, $T$ and $V$ leads to the exact result Eq. (\ref{g1}).\cite%
{ogu,hbo} Here we calculate numerically $\widetilde{\Sigma }^{r}$ up to
order $\widetilde{u}^{2}$, for finite $kT$ and $eV$ but smaller or of the
order of the Kondo energy $kT_{K}$.

Ordinary PT up to second order in $U$ supplemented by an interpolative
perturbative approach (IPA), \cite{levy,kaju} (which corrects the
second-order result in order to reproduce exactly the atomic limit $U/\Delta
\rightarrow +\infty $) has been shown to describe well the conductance
through a quantum dot for $U\leq 8\Delta $.\cite{pro} The results agree with
those obtained using the finite temperature density matrix renormalization
group method.\cite{maru} Comparison of the spin dependent IPA \cite{pc,lady}
with exact diagonalization in finite systems shows very good agreement for $%
U=6.25\Delta $.\cite{pc} The extension of PT in $U^{2}$ to the
nonequilibrium case has been considered by Hershfield \textit{et al}.\cite%
{hersh} They found that for finite $V$, the current is conserved only in the
electron-hole symmetric Anderson model (SAM). Different self-consistent
approaches were proposed to overcome this shortcoming, by a suitable
election of the unperturbed Hamiltonian.\cite{none,levy} While these
approaches work well in absence of a magnetic field $B$, numerical
difficulties persist for small non-vanishing $B$ and $V$.\cite{none} 
%ACA
Here we will 
take $B=0$ and parameters corresponding to
the SAM for our numerical integrations. In this case the current is
conserved for each spin without the need to solve self-consistent equations.\cite{none}

We use the spin 1/2 Anderson model to describe a quantum dot interacting
with two conducting leads, one at the left and one at the right, with
chemical potentials $\mu _{L}$ and $\mu _{R}$ respectively, with $\mu
_{L}-\mu _{R}=eV$. We define the zero of energy by $\mu
_{L}=eV\Delta _{R}/\Delta $, where $\Delta =\Delta _{L}+\Delta _{R}$ 
and $\Delta _{\nu }=\pi \sum_{k}|V_{k\nu }|^{2}\delta (\omega
-\varepsilon _{eff}^{\sigma })$ (neglecting here the small \cite{none} 
dependence on $\omega $). The Hamiltonian is split into a
noninteracting part $H_{0}$ and a perturbation $H^{\prime }$ as
%
%\begin{eqnarray}
%H &=&H_{0}+H^{\prime },  \nonumber \\
%H_{0} &=&\sum_{k\alpha \sigma }\varepsilon _{k\alpha }c_{k\alpha \sigma
%}^{\dagger }c_{k\alpha \sigma }+\sum_{\sigma }\varepsilon _{eff}^{\sigma
%}n_{d\sigma } \nonumber \\
%+\sum_{k\alpha \sigma }(V_{k\alpha }c_{k\alpha \sigma
%}^{\dagger }d_{\sigma }+\text{H.c.}),  \nonumber \\
%H^{\prime } &=&(E_{d}-\varepsilon _{eff}^{\sigma })\sum_{\sigma }n_{d\sigma
%}+Un_{d\uparrow }n_{d\downarrow },  \label{h}
%\end{eqnarray}
\begin{equation}
\left. \begin{aligned} H &=H_{0}+H^{\prime },\\ H_{0} &=\sum_{k\nu \sigma
}\varepsilon _{k\nu }\,c_{k\nu \sigma }^{\dagger }c_{k\nu \sigma
}+\sum_{\sigma }\varepsilon _{eff}^{\sigma }\,n_{d\sigma }\\ &+\sum_{k\nu
\sigma }\left(V_{k\nu }\, c^{\dagger }_{k\nu\sigma}d_{\sigma
}+\text{H.c.}\right),\\ H^{\prime } &=\sum_{\sigma }\left(E_{d}-\varepsilon
_{eff}^{\sigma }\right)\,n_{d\sigma}+U\,n_{d\uparrow }n_{d\downarrow },
\end{aligned}\right.
\end{equation}
where $\nu =L,R$ refers to the left and right leads. In general $\varepsilon
_{eff}^{\sigma }$ should be determined selfconsistently, but for the SAM
with $B=0$, $\varepsilon _{eff}^{\sigma }=0$.\cite{none} We obtain the
conductance $G=dI/dV$ from numerical differentiation of the current $I$,
which can be written as \cite{meir}
\begin{equation}
I=\frac{2e}{h}\int d\omega A\pi \Delta \rho (\omega )[f_{L}(\omega
)-f_{R}(\omega )],  \label{i}
\end{equation}
where $f_{\nu }(\omega )=f(\omega -\mu _{\nu })$, $f(\omega )=1/(e^{\omega
/kT}+1)$, $A=4\Delta _{L}\Delta _{R}/\Delta ^{2}$ indicates the degree of
asymmetry of the hybridization of the dot with both leads, and $\rho (\omega
)=-\text{Im}G_{d\sigma }^{r}(\omega )/\pi $ where $G_{d\sigma }^{r}(\omega )$
is the retarded Green's function of the electrons at the dot for spin $%
\sigma $, which can be written as \cite{none}
\begin{equation}
G_{d\sigma }^{r}(\omega )=\frac{1}{\omega -\varepsilon _{eff}^{\sigma
}+i\Delta -\Sigma _{\sigma }^{r}(\omega )}.  \label{gr}
\end{equation}

Within RPT, the low frequency part of $G_{d\sigma }^{r}(\omega )$ can be
approximated as \cite{he1}
\begin{equation}
\widetilde{G}_{d\sigma }^{r}(\omega )\simeq \frac{z}{\omega -\widetilde{%
\varepsilon}_{eff}^{\sigma }+i\widetilde{\Delta}-\widetilde{\Sigma}_{\sigma
}^{\text{rem}}(\omega )},  \label{gra}
\end{equation}
where $z=[1-\partial \Sigma _{\sigma }^{r}/\partial \omega ]^{-1}$, $%
\widetilde{\varepsilon}_{eff}^{\sigma }=z[\varepsilon _{eff}^{\sigma
}+\Sigma _{\sigma }^{r}(0)]$, $\widetilde{\Delta}=z\Delta $, and $\widetilde{%
\Sigma}_{\sigma }^{\text{rem}}(\omega )=z\Sigma _{\sigma }^{\text{rem}%
}(\omega )$, where the remainder retarded self-energy is defined as
\begin{equation}
\Sigma _{\sigma }^{\text{rem}}(\omega )=\Sigma _{\sigma }^{r}(\omega
)-\Sigma _{\sigma }^{r}(0)-\omega \partial \Sigma _{\sigma }^{r}/\partial
\omega .  \label{rem}
\end{equation}

In Eqs. (\ref{gra}) and (\ref{rem}), $\Sigma _{\sigma }^{r}(0)$ and $%
\partial \Sigma _{\sigma }^{r}/\partial \omega $ are evaluated at $\omega
=T=V=0$. A comparison between $G_{d\sigma }^{r}(\omega )$ (calculated within
PT) and $\widetilde{G}_{d\sigma }^{r}(\omega )$ with 
$\widetilde{\Sigma }_{\sigma }^{r}(\omega )=0$, for a case with non-trivial frequency dependent 
$\Delta (\omega )$ is provided in Ref. \onlinecite{vaug}, showing a very good
agreement for low $|\omega |$. For large values of $U/\Delta $, ordinary PT
in $U$ is not reliable and $z$ is in principle not known, although it can be
obtained from exact Bethe ansatz calculations. However, replacing $\Delta $
by $\widetilde{\Delta }/z$ in Eq. (\ref{i}), $z$ cancels and the current is
expressed in terms of renormalized parameters 
$\widetilde{\varepsilon }_{eff}^{\sigma }$, $\widetilde{\Delta }$, and $\widetilde{\Sigma }_{\sigma
}^{\text{rem}}(\omega )$. In the SC limit $U/\Delta \rightarrow \infty $,
Hewson \cite{he1} has shown that the ratio of renormalized parameters is $%
\widetilde{u}=\widetilde{U}/(\pi \widetilde{\Delta })=1$ and furthermore,
defining the Kondo temperature $T_{K}^{C}$ by the linear term in the
specific heat in this limit $\gamma_C =\pi ^{2}k/(6T_{K}^{C})$, one obtains
%\begin{equation}
$\widetilde{\Delta }=4kT_{K}^{C}/\pi $.  
%\label{der}
%\end{equation}
In the SAM at $B=0$, which we shall use, $\Delta _{L}=\Delta _{R}$, $%
E_{d}=-U/2$, and this implies $\widetilde{\varepsilon }_{eff}^{\sigma }=0$.\cite{vaug} 
Experimentally, $\Delta _{L}\sim \Delta _{R}$
and the scaling results do not depend on the gate voltage, which controls 
$E_{d}$. Furthermore, the value of $E_{d}$ is irrelevant
in the SC limit. This justifies the use of the SAM. Finally we use
\begin{equation}
\widetilde{\Sigma }_{\sigma }^{\text{rem}}(\omega )=\widetilde{\Sigma }%
_{\sigma }^{r}(\omega )-\widetilde{\Sigma }_{\sigma }^{r}(0)-\omega \partial 
\widetilde{\Sigma }_{\sigma }^{r}/\partial \omega ,  \label{remre}
\end{equation}
where $\widetilde{\Sigma }_{\sigma }^{r}(\omega )$ is obtained using
nonequilibrium PT up to second order in $\widetilde{u}$
%=\widetilde{U}/(\pi \widetilde{\Delta })$. 
Details of the
nonequilibrium RPT were published in previous works \cite{ogu,hbo}. The
second-order diagram has two sums over Matsubara frequencies. For $%
\widetilde{\Delta }$ independent of frequency, one of the sums can be done
analytically, which simplifies the numerical evaluation. Explicit expressions for
the retarded and lesser self-energies are given in the appendix of Ref. %
\onlinecite{none}. It is known \cite{yam} that $\partial \widetilde{\Sigma }%
_{\sigma }^{r}/\partial \omega =-(3-\pi ^{2}/4)\widetilde{u}^{2}$.

\begin{figure}[tbp]
\includegraphics[width=8cm]{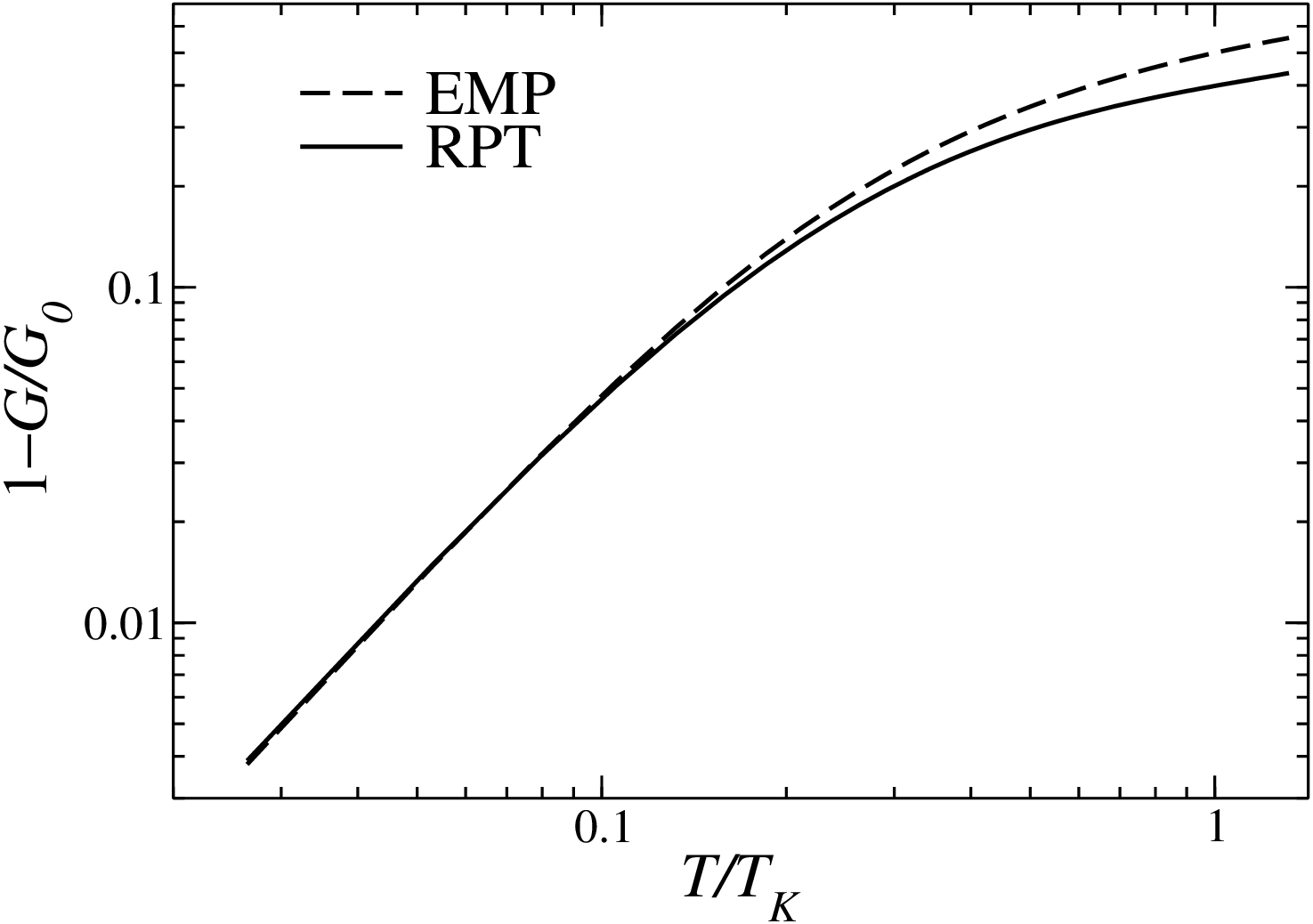}
\caption{Conductance shift for $V=0$ as a function of temperature for $%
\widetilde{u}=1$. Dashed line corresponds to the empirical (EMP) curve Eq. (%
\protect\ref{ge}).}
\label{t1}
\end{figure}

We begin by analyzing the case $\widetilde{u}=1$, which corresponds to the
strong coupling (SC) limit $U\rightarrow \infty $. As in the experiment,
\cite{grobis} we obtain $T_{K}$ by a fit of the temperature dependence of the
conductance for $V=0$ to Eq. (\ref{ge}) for 
$T/T_{K}<0.25$. In Fig.~\ref{t1} we compare Eq. (\ref{ge}) with our result.
The fit is very good for low $T/T_{K}$, while for $T/T_{K}\simeq 1$, our
result lies below the empirical curve. Remarkably, this is also the case of
the experimental results.\cite{grobis} This deviation however is
outside the region of the fit and is irrelevant in the following discussion. 
From the fit we obtain $kT_{K}=0.757\widetilde{\Delta }$. The exact
results to order $T^{2}$ and $V^{2}$ can be written in the form \cite{ogu}
\begin{eqnarray}
\frac{G}{G_{0}} &\simeq &1-\frac{\pi ^{2}(1+2\widetilde{u}^{2})}{3}\left( 
\frac{kT}{\widetilde{\Delta}}\right) ^{2}  \notag \\
&&-\frac{4-3A+(2+3A)\widetilde{u}^{2}}{4}\left( \frac{eV}{\widetilde{\Delta}}
\right) ^{2}.  \label{ogu}
\end{eqnarray}

Note that in the Kondo limit $\widetilde{u} \rightarrow 1$, the coefficients
are independent of the asymmetry parameter $A$, in agreement with 
recent calculations with the Kondo model.\cite{sela} 
 
A comparison with the expansion of Eq. (\ref{ge}) (for $V=0$) up to second
order in $T$ leads to
\begin{equation}
\frac{kT_{K}}{\widetilde{\Delta}}=\frac{3s(2^{1/s}-1)}{\pi ^{2}(1+2%
\widetilde{u}^{2})},  \label{tkd}
\end{equation}
which for $\widetilde{u}=1$ implies 
$T_{K}=0.746\widetilde{\Delta}/k=0.949T_{K}^{C}$. The small discrepancy with the value $kT_{K}=0.757%
\widetilde{\Delta}$ obtained from the fit is due to the finite temperature
interval used for fitting.

\begin{figure}[tbp]
%[rbp]
\includegraphics[width=8cm]{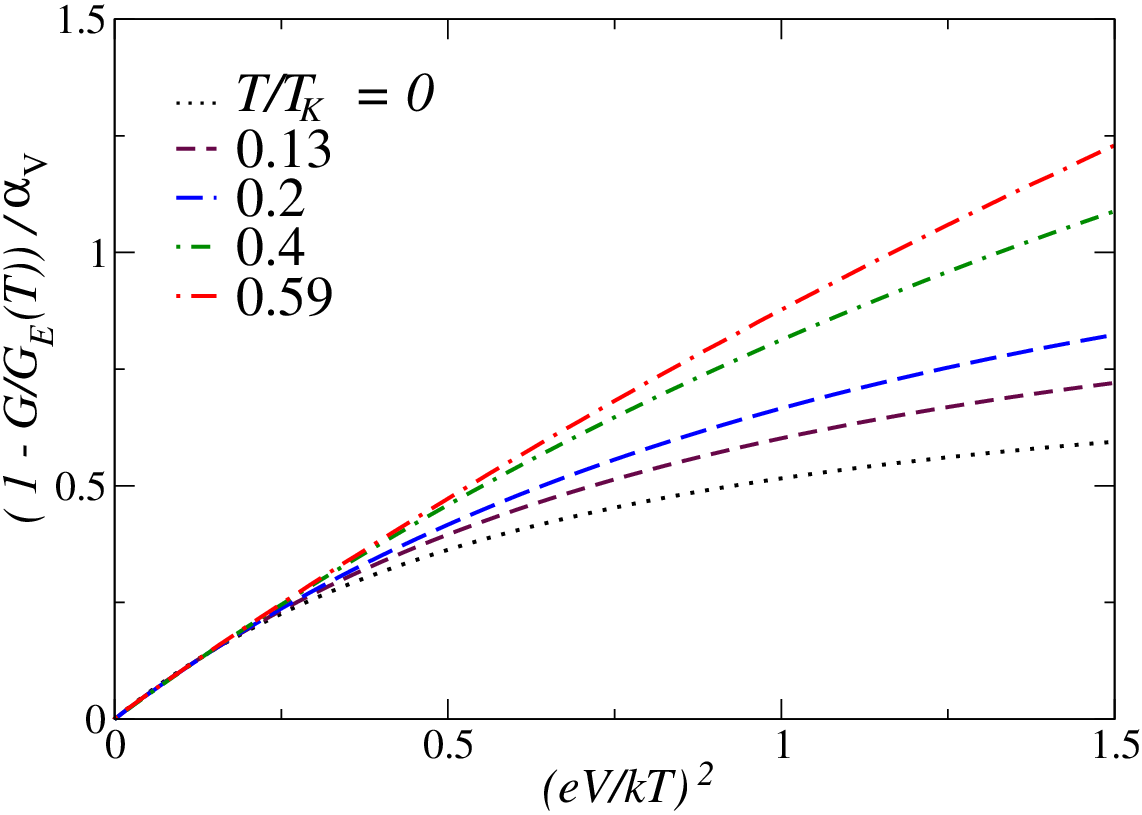}
\caption{(Color online) Scaled conductance shift as a function of bias
voltage for different temperatures and $\widetilde{u}=1$.}
\label{v1}
\end{figure}

Next we calculate the conductance $G=dI/dV$ for finite $T$ and $V$,  
by numerical differentiation of Eq. (\ref{i}) and compare the results with Eqs. 
(\ref{g2}) and (\ref{ge}). To obtain $\gamma $ we have fitted the current to a
polynomial with odd powers of $V$ up to $V^{3}$within the range $0\leq
eV/kT_{K}\leq 0.4$, as in the experiments. Inclusion of terms of
order $V^{5}$ practically does not modify the results. The
resulting shift in the conductance $(1-G/G_{E}(T))/{\alpha }_{V}$ scaled as
in the experimental work with
${\alpha }_{V}=\alpha c_{T}/[1+(\gamma /\alpha -1)c_{T}(T/T_{K})^{2}],$
%\begin{equation}
%{\alpha }_{V}=\frac{\alpha c_{T}}{1+(\gamma /\alpha -1)c_{T}(T/T_{K})^{2}},
%\label{at}
%\end{equation}
is shown in Fig.~\ref{v1}. 

%ACA
From the fit for $T=0$ we
obtain $\alpha =0.151$.
This agrees with Eq. (\ref{ogu}), which predicts 
a ratio of the coefficients of the $V$ and $T$ dependence of the conductance
[see Eq. (\ref{g1})]  
$\alpha =3/(2\pi ^{2})=0.152$ for $\widetilde{u}=1$, independently of the asymmetry $A$.  .  
The small discrepancy is probably due to numerical
errors in the integration near the singularities of the integrand.\cite{none}
The 
resulting values of $\gamma$ are near 0.75 for $0.13 \leq T/T_K \leq 0.4$ and increase to
0.91 for $T/T_K=0.59$ and 1.2 for $T/T_K=0.79$ (not shown). These values
are larger 
larger than the experimentally reported $\gamma =0.5\pm 0.1$. In addition,
while the quadratic scaling with $V$ was expected, for most cases
the observed exponent $\alpha =0.10\pm 0.015$ is smaller than the value 
$\alpha _{\text{SC}}^{\text{SAM}}=3/(2\pi ^{2})$ of the SAM in the SC limit 
$U\rightarrow \infty $. Note however that for some of the measured systems 
$\alpha $ approaches this value (Fig. 3 of Ref. \onlinecite{grobis} for 
$V_{G}\sim -0.195$ mV). In addition, in comparison with other theoretical
predictions for $\alpha $, $3/\pi ^{2}$ (Ref. \onlinecite{schi}), 
$3/(8\pi^{2})$ (Ref. \onlinecite{kami}), and $4/\pi ^{2}$ (Ref. \onlinecite{konik}),
the above value of $\alpha $ lies closer to experiment. However, it
is clear that the Anderson model in the SC limit is not able to reproduce
quantitatively the experiment.

\begin{figure}[bp]
\vspace{0.4cm} \includegraphics[width=8cm]{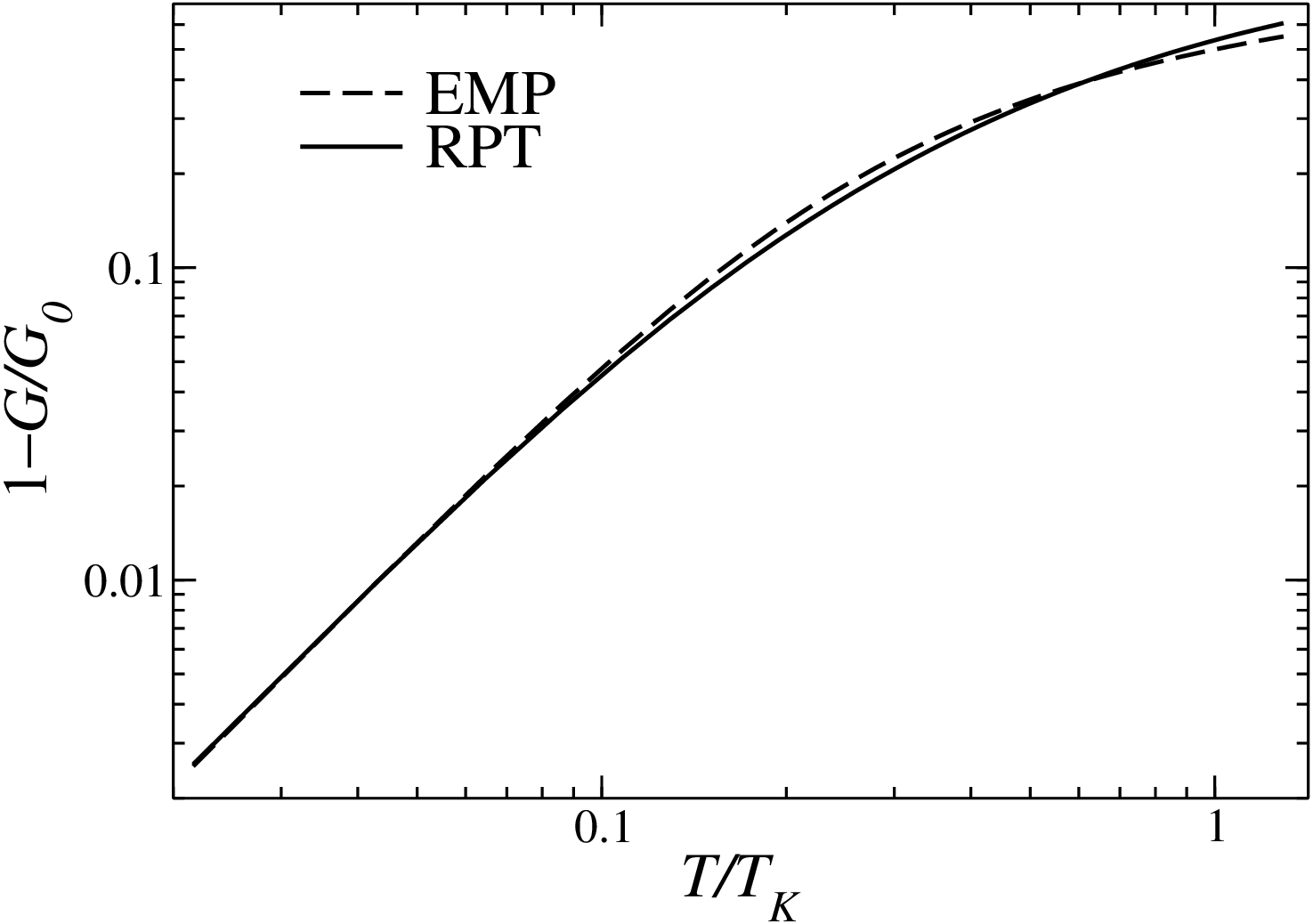}
\caption{Conductance shift for $V=0$ as a function of temperature for 
$\widetilde{u}=0.365$. Dashed line corresponds to Eq. (\protect\ref{ge}).}
\label{t2}
\end{figure}

A value of $\alpha <\alpha _{\text{SC}}^{\text{SAM}}$ can be
obtained
decreasing $\widetilde{u}$ [see Eq. (\ref{ogu})]
allowing some degree of intermediate valence.
We explore this possibility, within the SAM, allowing finite $U$
and therefore $\widetilde{u}<1$. This means that while the average
occupation of the dot is kept at the same value ($n=1$ in the SAM), some
charge fluctuations with the neighboring configurations is allowed. From
Eqs. (\ref{g1}) and (\ref{ogu}) one sees that $\alpha =0.1$ implies 
$\widetilde{u}=0.365$. We have repeated the calculations for this value of 
$\widetilde{u}$. The conductance as a function of temperature is displayed in
Fig.~\ref{t2}. We see that in this case, our result lies even
closer to the phenomenological curve, indicating that the effect of
temperature beyond $T^{2}$ is well described by our approximation
(which assumed $\widetilde{u}$ independent of $T$ and $V$). 
Proceeding with the fit
as in the experiment, we obtain $T_{K}=1.159\widetilde{\Delta }$, while Eq. 
(\ref{tkd}) gives $T_{K}=1.147\widetilde{\Delta }/k=1.461T_{K}^{C}$.

\begin{figure}[tp]
%\vspace{0.8cm}
\includegraphics[width=8cm]{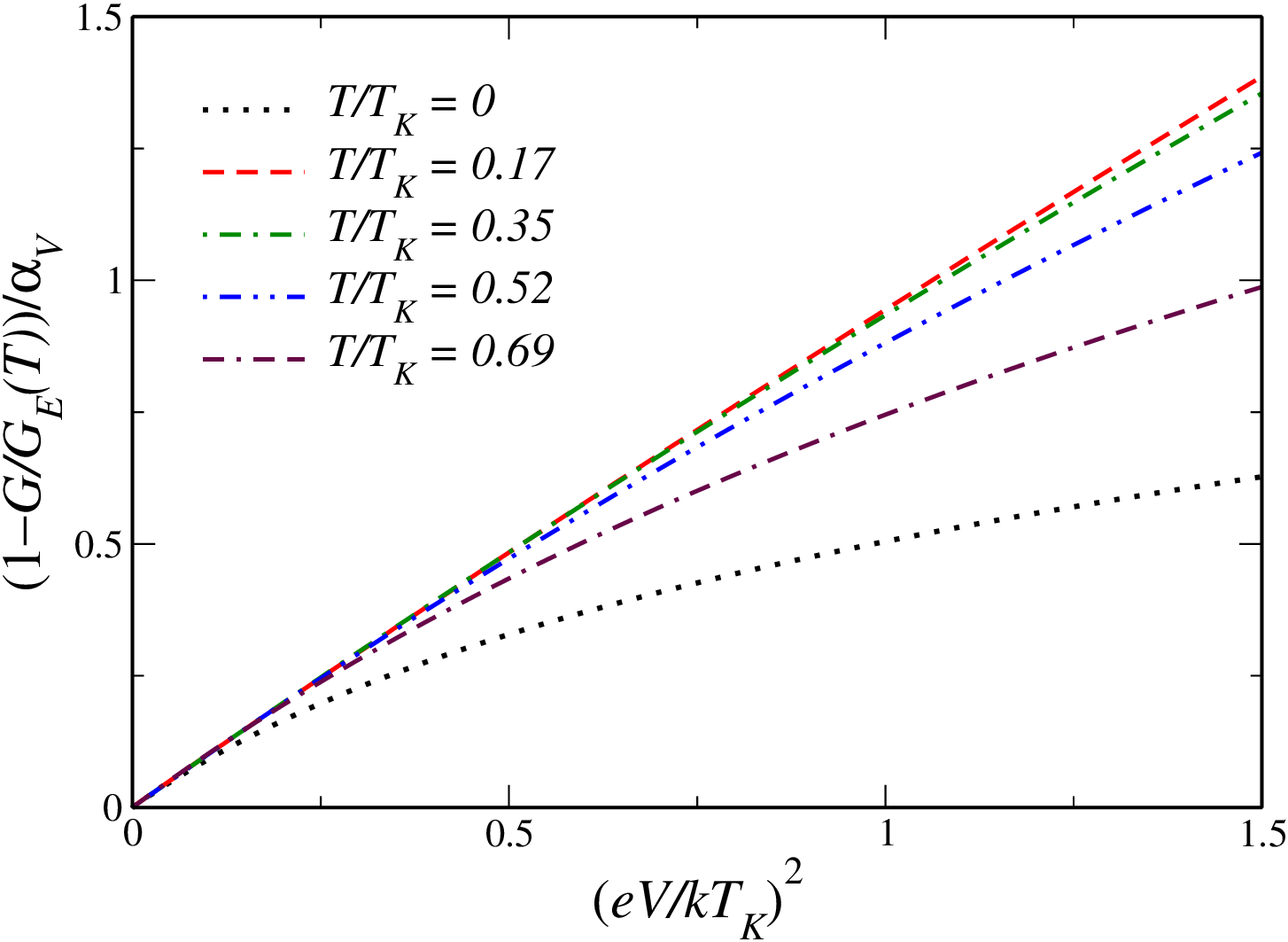}
\caption{(Color online) Scaled conductance shift as a function of bias
voltage for different temperatures and $\widetilde{u}=0.365$.}
\label{v2}
\end{figure}

In Fig.~\ref{v2} we show the scaled shift in the conductance $(1-G/G_{E}(T))/%
{\alpha}_{V}$ [see Eq. (\ref{ge})] for $\widetilde{u}=0.365$
as a function of the applied voltage for several temperatures. 

The corresponding values of
$\gamma$ are 0.75, 0.57, 0.49, 0.43  for $T/T_K$ =0.17, 0.35, 0.52, and 0.69  
respectively. Thus except for the smallest temperatures $T/T_K \leq 0.17$, in this case 
$\gamma$ agrees with the experimental value $\gamma =0.5\pm 0.1$.

The reader might wonder if different physical situations in which the Kondo
temperature can vary within a factor two are consistent with similar values
of the renormalized ratio 
$\widetilde{u}=\widetilde{U}/(\pi \widetilde{\Delta })$. 
In fact, while the Kondo temperature decreases exponentially
by increasing the ratio of the bare parameters $u=U/\pi \Delta $, $%
\widetilde{u}$ increases much slower,\cite{he1,ogu} being of course
$\widetilde{u}\sim u$ for small $u$ (including $u\sim 0.4$) and saturating at $\widetilde{u}=1$, 
for $u\rightarrow \infty $.

Finally, we note that if Eq. (1) with arbitrary exponents is used to fit the
conductance with finite voltage and temperature ranges,\cite{grobis} the
resulting exponents are slightly below 2 in agreement with experiment.\cite{grobis}

In summary, using nonequilibrium renormalized perturbation theory up to
second order in the renormalized perturbation parameter $\widetilde{u}$ for
the Anderson model in the symmetric case, we have examined the scaling
behaviour of the conductance, including terms beyond those quadratic in
temperature and bias voltage. In the strong coupling limit, the
model predicts an effect of voltage which is 50\% higher than observed and
the effects of terms of order $(VT)^{2}$ disagree with
experiment. If instead an important degree of valence fluctuations is
allowed, we obtain good agreement with recent experimental results.

The authors are supported by CONICET. This work was done in the framework of
projects PIP 5254 of CONICET and PICT 2006/483 of the ANPCyT.

\end{document}